\documentstyle[psfig]{mn}

\title[The Durham/UKST Galaxy Redshift Survey -- VII.]{The Durham/UKST Galaxy Redshift Survey -- VII. Redshift-space distortions in the power spectrum}
\author[P.~J.~Outram, Fiona~Hoyle, T.~Shanks]{P.~J.~Outram, Fiona~Hoyle, T.~Shanks\\
Department of Physics, Science Laboratories, South Road, Durham DH1 3LE.}

\begin{document}
\maketitle
\begin{abstract} 
We investigate the effect of redshift-space distortions in the power spectrum parallel and perpendicular to the observer's line of sight, $P^S(k_{\parallel},\mathbf{k}_{\perp})$, using the optically selected Durham/UKST Galaxy Redshift Survey. On small, non-linear scales anisotropy in the power-spectrum is dominated by the galaxy velocity dispersion; the `Finger of God' effect. On larger, linear scales coherent peculiar velocities due to the infall of galaxies into overdense regions are the main cause of anisotropy. According to gravitational instability theory these distortions depend only on the density and bias parameters via  $\beta \approx \Omega_m^{0.6}/b$. Geometrical distortions also occur if the wrong cosmology is assumed, although these would be relatively small given the low redshift of the survey.
To quantify these effects, we assume the real-space power spectrum of the APM Galaxy Survey, and fit a simple model for the redshift-space and geometrical distortions. Assuming a flat $\Omega_m = 1$ universe, we find values for the one-dimensional pairwise velocity dispersion of $\sigma_p = 410 \pm 170\;$km$\;$s$^{-1}$, and $\beta = 0.38 \pm 0.17$. An open $\Omega_m = 0.3$, and a flat $\Omega_m = 0.3$, $\Omega_{\Lambda} = 0.7$ universe yield $\sigma_p = 420\;$km$\;$s$^{-1}$, $\beta = 0.40$, and $\sigma_p = 440\;$km$\;$s$^{-1}$, $\beta = 0.45$ respectively, with comparable errors.   These results are consistent with estimates using the two-point galaxy correlation function, $\xi(\sigma,\pi)$, and favour either a low-density universe with $\Omega_m \sim 0.3$ if galaxies trace the underlying mass distribution, or a bias factor of $b \sim 2.5$ if $\Omega_m = 1$.
\end{abstract}

\section{Introduction}

When mapping the three-dimensional distribution of galaxies redshift is generally the easiest, and often the only distance estimator available. Redshifts, however, do not reflect real galaxy distances because of peculiar velocities relative to the Hubble flow. Whilst standard cosmologies assume that the Universe, and hence galaxy clustering, is isotropic in real-space, the effect of peculiar velocities and incorrect geometric assumptions can distort the picture in redshift-space. The anisotropy is dominated by the galaxy velocity dispersion in virialized clusters on the smallest scales. This is the `Finger of God' effect seen in redshift surveys. The main cause of anisotropy on larger, linear scales is the coherent peculiar velocity field of large-scale structure. The infall of galaxies into overdense regions makes their distribution appear slightly squashed in redshift-space, increasing the apparent overdensity. According to linear gravitational instability theory these distortions depend only on the density and bias parameters via the combination $\beta \approx \Omega_m^{0.6}/b$ (Kaiser 1987). Incorrect geometric assumptions through the assumed cosmological model also add an anisotropy on all scales. These redshift-space distortions can be exploited in order to constrain properties of the galaxy velocity field, and fundamental cosmological parameters such as $\Omega$ and $\Lambda$.

Redshift-space distortions can be investigated using the spatial two-point correlation function or its Fourier transform, the power spectrum.  In an ideal situation these two approaches are equivalent. When applied to finite, noisy data, however, they have different strengths, and should be viewed as complementary. In particular the power spectrum approach has the advantage that uncertainty in the mean galaxy density only affects the $k=0$ mode, whereas uncertainty in this quantity introduces errors in the correlation function measurement at all scales. Also, the power spectrum is more naturally related to theoretical model predictions, and the modelling of these anisotropies is simpler in redshift-space. 

Early attempts to investigate the anisotropies in the power spectrum (Cole, Fisher, \& Weinberg 1994) focussed on the linear regime, developing a simple estimator for $\beta$ by considering the quadrupole and monopole angular moments of the power spectrum. This model was improved to account for a small-scale velocity dispersion (Cole, Fisher, \& Weinberg 1995), but still ignored geometric distortions due to cosmological assumptions. Any geometric distortion has a similar effect as that of $\beta$, and if ignored this would lead to systematic errors. Whilst working at very low redshifts, where the geometric distortions are negligible, this approach is valid, but at the higher redshifts probed by the future large galaxy and quasar redshift surveys, cosmological effects need to be considered. Reducing the full two-dimensional information available in $P^S(k_{\parallel},\mathbf{k}_{\perp})$ to just the first two angular moments would make it impossible to discriminate the subtle angular differences between geometrical and redshift-space distortions. Ballinger, Peacock, \& Heavens (1996) developed a model for $P^S(k_{\parallel},\mathbf{k}_{\perp})$ taking all of these effects into account. 

In this paper we investigate the redshift-space distortions in the power spectrum parallel and perpendicular to the observer's line of sight, $P^S(k_{\parallel},\mathbf{k}_{\perp})$, using the optically selected Durham/UKST Galaxy Redshift Survey. We apply the model from Ballinger et al. (1996) to the power spectrum obtained and derive estimates for $\beta$, and the one-dimensional pairwise velocity dispersion, $\sigma_p$. Finally, we compare our results to those obtained from the same survey through analysis of the two-point correlation function (Ratcliffe et al. 1998a).

\section{Power Spectrum Analysis}

\subsection{Sample definition}
The power spectrum analysis is carried out on the volume limited sample, with $z \le z_{max}=0.06$, of the optically selected Durham/UKST Galaxy Redshift Survey, described by Hoyle et al. (1999). Full details of the construction of the survey can be found in Ratcliffe et al. (1998b). 
\subsection{Power spectrum estimation}

\begin{figure}
\centerline{\hbox{\psfig{figure=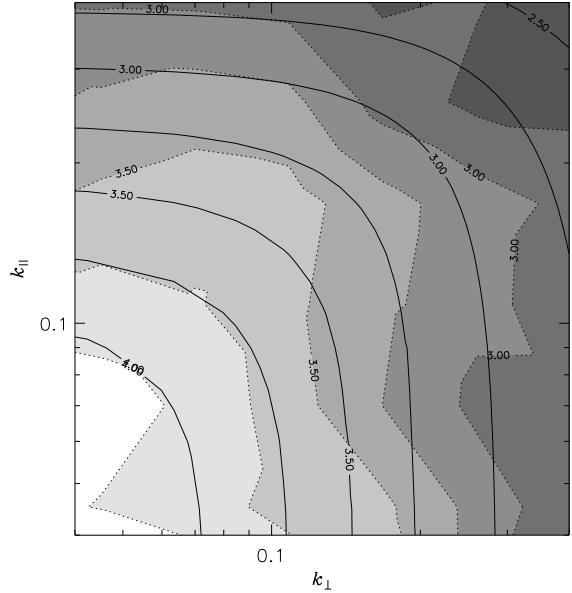,height=8.5cm}}}
\caption{$P^S(k_{\parallel},\mathbf{k}_{\perp})$ estimated from the Durham/UKST survey. Filled contours of constant ${\mathrm log} P$ are shown as a function of $k_{\parallel}$ and $\mathbf{k}_{\perp}$. Overlaid is the best-fit model for $P^S(k_{\parallel},\mathbf{k}_{\perp})$, assuming a flat $\Omega_m = 1$ universe, with $\beta = 0.38$ and $\sigma_p = 410\;$km$\;$s$^{-1}$, and the real-space APM power spectrum.
}
\label{fig2}
\end{figure}

The power spectrum is estimated using the method outlined in Hoyle et al. (1999) and Tadros \& Efstathiou (1996). The Durham/UKST Survey is divided into four regions in order to apply the distant observer approximation (see section~\ref{datageom}). Each region is embedded into a larger cubical volume, which is rotated such that the central line of sight lies along the same axis of the cube in each case. The density field is binned onto a 256$^3$ mesh, using nearest gridpoint assignment.
The power spectrum of each region is estimated using a Fast Fourier Transform, and the average of the resulting power spectra is taken. The result, binned logarithmically into $k_{\parallel}$ and $\mathbf{k}_{\perp}$ is plotted in figure~\ref{fig2}.

\subsection{Survey geometry}\label{datageom}
The Durham/UKST Survey covers a region of approximately $20^{\circ} \times 75^{\circ}$. In order to retain information about the line of sight, to calculate $P^S(k_{\parallel},\mathbf{k}_{\perp})$, the data have to be divided into subsamples that subtend a small solid angle on the sky, and the distant-observer approximation applied. The data are split into 4 regions each approximately $20^{\circ} \times 20^{\circ}$ during this analysis, and the average of the resulting power spectra is taken. 
The measured $P^S(k_{\parallel},\mathbf{k}_{\perp})$ is in fact a convolution between the true power spectrum and a function related to the sample window introduced due to the assumption of parallel lines of sight. This makes the measured power spectrum more isotropic, and hence causes a small, yet systematic underestimate of $\beta$. Although it is possible to correct for this effect (see Cole et al. 1994), the error in this approximation for a $20^{\circ} \times 20^{\circ}$ window is $< 1\%$ and considerably smaller than the statistical errors obtained in this analysis. Hence for the purposes of this analysis it is ignored. 

\begin{figure}
\centerline{\hbox{\psfig{figure=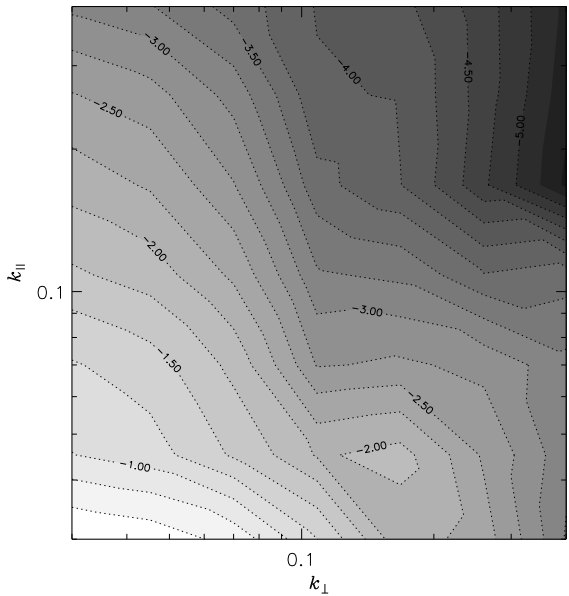,height=8.5cm}}}
\caption{The power spectrum of the window function. Filled contours of constant ${\mathrm log}(|\hat{W}(k_{\parallel},{\mathbf k}_{\perp})|^2)$ are shown as a function of $k_{\parallel}$ and $\mathbf{k}_{\perp}$.  For wavenumbers with $k_{\parallel} \ge0.06\;h\;$Mpc$^{-1}$ or $|\mathbf{k}_{\perp}|$ $ \ge0.1\;h\;$Mpc$^{-1}$, the window function power spectrum is a steep power law, varying as $\sim k^{-3}$ or steeper.
}
\label{fig1}
\end{figure}

The measured power spectrum is also convolved with the power spectrum of the window function. As we are considering a volume limited sample, each galaxy carries equal weight, and hence the survey window function, $W({\mathbf x})$, simply takes a value of unity in the volume of the universe included in the survey, and zero elsewhere. This is approximated using a catalogue containing a large number of unclustered points with the same radial and angular distribution as the survey, and its power spectrum is calculated in a similar manner to that of the data. Figure~\ref{fig1} shows the power spectrum of the window function, $|\hat{W}|^2$, as a function of $k_{\parallel}$ and $\mathbf{k}_{\perp}$.  

We can probe larger scales along the line of sight than across it; the depth of the survey is $\approx 175h^{-1}$Mpc, corresponding to $k_{\parallel min}\approx 0.04\;h\;$Mpc$^{-1}$, whereas the maximum scale probed at this depth across the line of sight in a  $20^{\circ} \times 20^{\circ}$ region is $\approx 85h^{-1}$Mpc, corresponding to $|\mathbf{k}_{\perp min}|$ $ \approx 0.07\;h\;$Mpc$^{-1}$. The effect of this can be seen in the anisotropy of $|\hat{W}|^2$ on the largest scales. Power on the largest scales across the line of sight is lost when the data are split into sub-samples. The choice of 4 sub-samples is a compromise between minimizing the effects of the distant-observer approximation and maximising the volume of ${\mathbf k}$-space we can probe.  

The window function is sharply peaked in ${\mathbf k}$-space, varying as a steep power law,  $\sim k^{-3}$ or steeper; much steeper than the expected galaxy power spectrum. Therefore the convolution should have little effect at wavenumbers where $|\hat{W}|^2$ takes a value less than about a tenth of its maximum value.  At wavenumber separations where  $|\hat{W}|^2$ takes a value greater than this, the convolution will have an effect on our estimate of the galaxy power spectrum. To avoid this, we consider only those regions with $k_{\parallel} \ge0.06\;h\;$Mpc$^{-1}$ or $|\mathbf{k}_{\perp}|$ $ \ge0.1\;h\;$Mpc$^{-1}$.

\subsection{Mock catalogues}

In order to test our power spectrum estimator, and to estimate the statistical errors on the recovered power spectrum, mock catalogues are used. A total of 40 mock catalogues are constructed using the \emph{Hubble Volume} $\tau$CDM simulation from the Virgo Consortium (Colberg et al. 1998). They have the same radial and angular selection as the Durham/UKST Survey, and approximately the same clustering amplitude.  For further details on the construction of the mock catalogues see Hoyle et al. (1999).
The power spectrum of each catalogue was measured in exactly the same way as for the data.
\begin{figure}
\centerline{\hbox{\psfig{figure=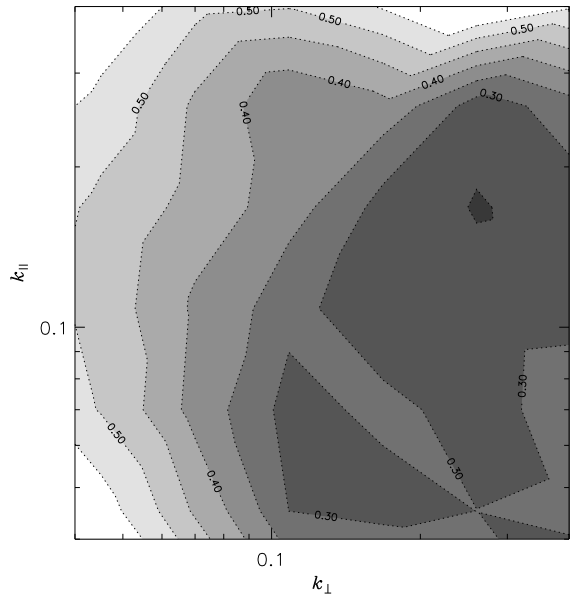,height=8.5cm}}}
\caption{The fractional errors in $P^S(k_{\parallel},\mathbf{k}_{\perp})$, as estimated using mock catalogues, are shown as a function of $k_{\parallel}$ and $\mathbf{k}_{\perp}$. 
}
\label{fig4}
\end{figure}

The fractional errors, calculated from variance in the power spectrum measurements across the 40 mock catalogues, are displayed in figure~\ref{fig4}. The errors on the power spectrum estimate are lowest at high values of $|\mathbf{k}_{\perp}|$. This is because $\mathbf{k}_{\perp}$ represents two dimensions in Fourier space, and hence
the Fourier space volume sampled in any bin $\propto \Delta k_{\parallel}$ $\times |\mathbf{k}_{\perp}|$ $  \Delta |\mathbf{k}_{\perp}|$.

\begin{figure}
\centerline{\hbox{\psfig{figure=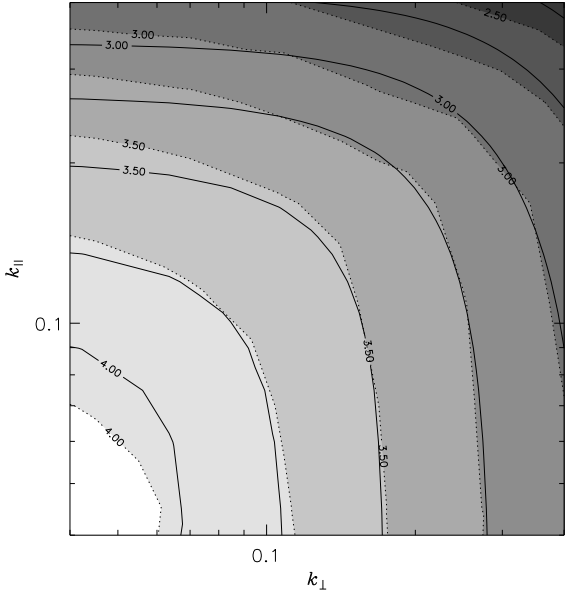,height=8.5cm}}}
\caption{The average $P^S_{mock}(k_{\parallel},\mathbf{k}_{\perp})$ estimated from 40 mock catalogues, constructed from the Virgo Consortium's \emph{Hubble Volume} $\tau$CDM simulation. Filled contours of constant ${\mathrm log} P$ are shown as a function of $k_{\parallel}$ and $\mathbf{k}_{\perp}$. Overlaid is the best-fit model for $P^S_{mock}(k_{\parallel},\mathbf{k}_{\perp})$, assuming a flat $\Omega_m = 1$ universe, with $\beta = 0.37$ and $\sigma_p = 540\;$km$\;$s$^{-1}$.
}
\label{figmock}
\end{figure}

As well as using the mock catalogues to estimate the statistical errors, we can also investigate possible systematics, such as the effect of the window function, by fitting the redshift-space distortions in the mock power spectra, using the method described in the next section. The average of the 40 mock catalogue power spectra is shown in figure~\ref{figmock}, together with the best fitting model. The errors are considerably smaller than those of the data, as this is the average over 40 realizations.
 The discrepancy on the largest scales is partly due to the effects of the window function. To minimise the effects of this uncertainty, the region with $k_{\parallel} \ge0.06\;h\;$Mpc$^{-1}$ and $|\mathbf{k}_{\perp}|$ $ \ge0.1\;h\;$Mpc$^{-1}$ is excluded from the fit. Also, fractional errors decrease with increasing 
$|\mathbf{k}_{\perp}|$, as shown in figure~\ref{fig4}, and so the agreement between the mock catalogues and the model is best at large $|\mathbf{k}_{\perp}|$.
The best-fit model for $P^S_{mock}(k_{\parallel},\mathbf{k}_{\perp})$, assuming a flat $\Omega_m = 1$ universe, has $\beta = 0.37$ and $\sigma_p = 540\;$km$\;$s$^{-1}$. The pairwise velocity measurement is in remarkable agreement with that found in the simulation; Jenkins et al. (1998) measured $\sigma_p \approx 530\;$km$\;$s$^{-1}$  for the whole simulation on large scales. The value of $\beta$ depends on the precise bias prescription used, but is in good agreement with other estimates. Hence we conclude that our method produces a robust measurement of the redshift-space distortions, and any systematic uncertainties in our measurements, especially due to the effects of window function, are small on the scales considered.

\section{Modelling Redshift Distortions}
The model that we fit to the data is based on that presented in Ballinger, Peacock and Heavens (1996). For clarity and convenience we review the main details below.
\subsection{Geometrical distortions}

The power spectrum analysis was carried out assuming an EdS universe. If the true geometry is different then our distance calculations will be wrong by a factor $f_{\perp}$ perpendicular to the line of sight, and $f_{\parallel}$ along the line of sight (as defined in Ballinger et al. 1996). Thus we can define a geometric flattening factor:

\begin{equation}\label{flateqn}
F(z)=\frac{f_{\parallel}}{f_{\perp}}=1+\frac{1}{4}(1-\Omega_m+2\Omega_{\Lambda})z + O(z^2)
\end{equation}
(Ballinger et al. 1996). For an open ($\Omega_m = 0.3$, $\Omega_{\Lambda} = 0.0$), and lambda-dominated ($\Omega_m = 0.3$, $\Omega_{\Lambda} = 0.7$) universe, $F \approx 1.01$ and $1.03$ respectively at the survey limit. As the depth of the survey is increased this effect becomes much more important. The effect this has on the power spectrum is given by
\begin{equation}\label{geomeqn}
P_{assumed}(k_{\parallel},{\mathbf k}_{\perp})= 
\frac{P_{true}(k)}{f_{\perp}^{3+n}F}  
\left[1+\mu^2\left(\frac{1}{F^2}-1\right)\right]^{\frac{n}{2}}
\end{equation}
(Ballinger et al. 1996) where $\mu=k_{\parallel}/k$, and $n$ is the spectral index of the power spectrum. 
\subsection{Redshift distortions}
On large, linear scales coherent peculiar velocities due to the infall of galaxies into overdense regions are the main cause of anisotropy. Kaiser (1987) showed that this anisotropy takes a very simple form in redshift-space, depending only on the density and bias parameters via the combination $\beta \approx \Omega_m^{0.6}/b$:
\begin{equation}\label{betaeqn}
P^S(k_{\parallel},{\mathbf k}_{\perp}) =
P^R(k)\left[1+\beta\mu^2\right]^2
\end{equation}
where $P^S$, and $P^R$ refer to the redshift-space and real-space power spectra respectively. The Kaiser formula assumes the distant-observer approximation, and hence is valid in this case due to the method of estimation of $P^S(k_{\parallel},{\mathbf k}_{\perp})$ (\S~\ref{datageom}).

On small, non-linear scales anisotropy in the power-spectrum is dominated by the galaxy velocity dispersions in virialized clusters. This is modelled by introducing a damping term. Assuming an exponential line of sight pairwise velocity ($\sigma_p$)\footnote{In power spectra $\sigma_p$ is implicitly divided by $H_0$ and quoted in units $h^{-1}\;$Mpc. ($H_0=100h\;$km$\;$s$^{-1}\;$Mpc$^{-1}$) } distribution (Davis \& Peebles 1983), this is given by a Lorentzian factor in redshift-space:
\begin{equation}
D\left[k\mu\sigma_p\right]=\frac{1}{1+\frac{1}{2}\left(k\mu\sigma_p\right)^2}
\end{equation}
This model is only a good approximation in mildly non-linear situations. Cole et al. (1995) used numerical simulations to show that it is effective for wavelengths $\lambda \ga 15h^{-1}\;$Mpc, or $k\la0.4$. At such wavenumbers assuming a Gaussian velocity dispersion has a negligible difference on the resulting model (Ballinger et al. 1996).

Combining these effects leads to the final model:
\begin{eqnarray}\label{model}
P^S(k_{\parallel},{\mathbf k}_{\perp}) &= &
\frac{P^R(k)}{f_{\perp}^{3+n}F}  
\left[1+\mu^2\left(\frac{1}{F^2}-1\right)\right]^{\frac{n-4}{2}}\nonumber\\
&\times& \left[1+\mu^2\left(\frac{\beta+1}{F^2}-1\right)\right]^2
D\left[k\mu\sigma_p^{\prime}\right]
\end{eqnarray}
(Ballinger et al. 1996) where $\sigma_p^{\prime}=\sigma_p/f_{\parallel}$. There are several free parameters in the model; $F, \beta, \sigma_p$, and $P^R(k)$, the underlying real-space power spectrum. To reduce the uncertainties on the parameters of interest we assume the real-space power spectrum of the APM Survey galaxies (Baugh \& Efstathiou 1993). When fitting to the $\tau$CDM mock catalogues, we assume instead a $\tau$CDM real-space power spectrum.

\subsection{$\chi^2$ fitting}

We fit the model to the data using a $\chi^2$ technique:
\begin{equation}
\chi^2=\sum\frac{(P_{obs}- P_{mod})^2}{\sigma_{ P(k)}^2}
\end{equation}
The errors, $\sigma_{P(k)}$, are those estimated from the mock catalogue power spectra. 
Only those wavenumbers with $k_{\parallel}>0.06$ or $ |{\mathbf k}_{\perp}|>0.1$, and $k<0.4$ are used in the fit. The former constraints are applied to remove the effects of the window function, and the latter to prevent excessive non-linearity, where the model breaks down.

An underlying cosmology is assumed and the $\beta$ and $\sigma_p$  parameters are adjusted until a minimum value of $\chi^2$ is obtained. The errors are calculated using Monte Carlo simulations of data sets drawn from the best fitting model, using the mock catalogue errors. The results are displayed in table~\ref{table}. 

\begin{table}
\caption{Minimum $\chi^2$ fits (over 23 bins) for $\sigma_p$ and $\beta$ to $P^S(k_{\parallel},\mathbf{k}_{\perp})$ estimated from the Durham/UKST survey. The open and lambda cosmological models  both assume $\Omega_m = 0.3$, and $\Omega_{\Lambda} = 0.0$ and 0.7 respectively.}
\label{table}
\begin{tabular}{@{}llcc}
Cosmology& $\sigma_p$/km$\;$s$^{-1}$&$\beta$&$\chi^2$\\
\hline
EdS&410$\pm$170&0.38$\pm$0.17 &14.18\\
Open&420$\pm$150&0.40$\pm$0.19 &14.11\\
$\Lambda$&440$\pm$200&0.45$\pm$0.19 &13.78\\
\hline
\end{tabular}
\end{table}
\begin{figure}
\centerline{\hbox{\psfig{figure=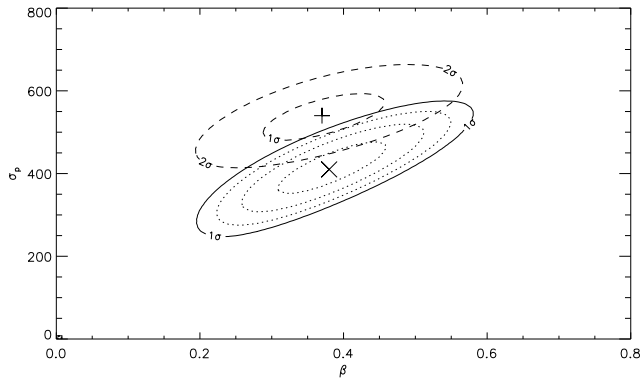,width=9cm}}}
\caption{Contours of increasing $\chi^2$ in the $\sigma_p$ -- $\beta$ plane, assuming a flat $\Omega_m = 1$ universe. The solid contour represents the 1-$\sigma$ error in the Durham/UKST survey estimate, the $\chi^2$ value of which was determined separately using Monte Carlo simulations.
The best fit values obtained ($\times$) are a one-dimensional pairwise velocity dispersion of $\sigma_p = 410 \pm 170\;$km$\;$s$^{-1}$, and $\beta = 0.38 \pm 0.17$. The 1-$\sigma$ and 2-$\sigma$ error countours are also shown for the fit to the average mock catalogue power spectrum (dashed lines), together with the best fit values ($+$) of $\sigma_p = 540\;$km$\;$s$^{-1}$, and $\beta = 0.37$.
}
\label{fig3}
\end{figure}

\section{Discussion}

Assuming an EdS universe, we find values for the one-dimensional pairwise velocity dispersion of $\sigma_p = 410 \pm 170\;$km$\;$s$^{-1}$, and $\beta = 0.38 \pm 0.17$. An open or lambda-dominated universe yield similar results (Table~\ref{table}).  $\chi^2$ contours for the fit are shown in figure~\ref{fig3}. Although the quoted errors assume independence of the two parameters, they are clearly related, and a tighter joint constraint would be possible. 

It is possible that the  assumed $P^R(k)$ in the model may differ from the true $P^R(k)$ of the Durham/UKST Galaxy Survey. Any differences could introduce systematic errors to the measurements of $\sigma_p$ and  $\beta$. To investigate this we repeated the fits described above, simultaneously fitting the amplitude of the real-space power spectrum, whilst still assuming the shape of the APM $P^R(k)$.  The additional parameter slightly increased uncertainty in the fitted values of $\sigma_p$ and  $\beta$, and led to a marginally higher value for the latter ($\beta=0.51\pm0.24$ in an EdS universe). The fitted amplitude, however, was found to be consistent with that of the original APM measurement, and we conclude that the assumption of the APM $P^R(k)$ amplitude and shape is adequate for the purposes of this paper. However, in a more extensive study, using large redshift survey datasets such as 2dF or Sloan, it may be necessary to fit the real-space power spectrum self-consistently within the model to minimize any systematic errors.

These results are consistent with other estimates from the Durham/UKST Galaxy Redshift Survey.  Ratcliffe et al. (1998a) determined values of $\beta=0.48$ and $\sigma_p = 416$ using the two-point galaxy correlation function, $\xi(\sigma,\pi)$. Hoyle et al. (1999) compared the Durham/UKST redshift-space power spectrum to the real-space APM power spectrum in order to estimate the redshift distortions, deriving slightly higher values of $\beta = 0.60 \pm 0.35$ and $\sigma_p \approx 450 \pm 200\;$km$\;$s$^{-1}$ (Hoyle et al. actually estimated the one-particle velocity dispersion $\sim \sigma_p/\sqrt{2}$). The $\beta$ estimate favours either a low-density universe with $\Omega_m \sim 0.3$ if galaxies trace the underlying mass distribution, or a bias factor of $b \sim 2.5$ if $\Omega_m = 1$.

\section*{Acknowledgements}

\noindent We would like to thank Carlton Baugh for helpful discussions. We acknowledge the help of Alison Broadbent, Chris Collins, Richard Fong, Anthony Oates, Quentin Parker, Andrew Ratcliffe, and Fred Watson in the construction of the Durham/UKST survey. FH acknowledges the support of a PPARC studentship.

\end{document}